\documentclass[
journal=nalefd, 
manuscript=letter]{achemso}
\pdfoutput=1 

\usepackage[version=3]{mhchem} 
\usepackage{enumitem}
\usepackage{graphicx}
\usepackage{mathrsfs}
\usepackage{makeidx}         
\usepackage{graphicx}        
\usepackage{gensymb}

\usepackage{bm}
\usepackage{amsmath}
\usepackage{amssymb}
\usepackage{color}
\usepackage[hidelinks]{hyperref}
\usepackage[framemethod=TikZ]{mdframed}

\SetLabelAlign{Center}{\hfil#1\hfil}

\mdfdefinestyle{BenFrame}{%
    outerlinewidth=0pt,
    roundcorner=10pt,
    innertopmargin=\baselineskip,
    innerbottommargin=\baselineskip,
    innerrightmargin=20pt,
    innerleftmargin=20pt,
    backgroundcolor=gray!10!white}

\author{Yongsop~Hwang}
\affiliation[SZU]
{Nanophotonics Research Centre, Shenzhen University \& Key Laboratory of Optoelectronic Devices and Systems of Ministry of Education and Guangdong Province, College of Optoelectronic Engineering, Shenzhen University, Shenzhen 518060, China}
\alsoaffiliation[RMIT]
{School of Engineering, RMIT University, Melbourne, VIC 3001, Australia}
\author{Ben~Hopkins}
\affiliation[ANU]
{Research School of Physics and Engineering, Australian National University, Canberra, ACT 2601, Australia}
\alsoaffiliation[AEP]
{School of Applied and Engineering Physics, Cornell University, Ithaca, NY 14853, USA}
\author{Dapeng~Wang}
\affiliation[SZU]
{Nanophotonics Research Centre, Shenzhen University \& Key Laboratory of Optoelectronic Devices and Systems of Ministry of Education and Guangdong Province, College of Optoelectronic Engineering, Shenzhen University, Shenzhen 518060, China}
\author{Arnan~Mitchell}
\affiliation[RMIT]
{School of Engineering, RMIT University, Melbourne, VIC 3001, Australia}
\author{Timothy~J.~Davis}
\affiliation[Melb]
{School of Physics, University of Melbourne, VIC 3052, Australia}
\author{Jiao~Lin}
\affiliation[SZU]
{Nanophotonics Research Centre, Shenzhen University \& Key Laboratory of Optoelectronic Devices and Systems of Ministry of Education and Guangdong Province, College of Optoelectronic Engineering, Shenzhen University, Shenzhen 518060, China}
\alsoaffiliation[RMIT]
{School of Engineering, RMIT University, Melbourne, VIC 3001, Australia}
\email{jiao.lin@rmit.edu.au}
\author{Xiao-Cong~Yuan}
\affiliation[SZU]
{Nanophotonics Research Centre, Shenzhen University \& Key Laboratory of Optoelectronic Devices and Systems of Ministry of Education and Guangdong Province, College of Optoelectronic Engineering, Shenzhen University, Shenzhen 518060, China}
\email{xcyuan@szu.edu.cn}

\title{Optical chirality from dark-field illumination of planar plasmonic nanostructures}

\begin{document}

\newpage
\begin{abstract}
Dark-field illumination is shown to make {\sl planar chiral} nanoparticle arrangements exhibit circular dichroism in extinction analogous to true chiral scatterers.
Circular dichrosim is experimentally observed at the maximum scattering of single oligomers consisting rotationally symmetric arrangements of gold nanorods, with strong agreement to numerical simulation. 
A dipole model is developed to show that this effect is caused by a difference in the geometric projection of a nanorod onto the handed orientation of electric fields created by a circularly polarized dark-field that is normally incident on a glass substrate. 
Owing to this geometric origin, the wavelength of the peak chiral response is also experimentally shown to shift depending on the separation between nanoparticles. 
All presented oligomers have physical dimensions less than the operating wavelength, and the applicable extension to closely packed planar arrays of oligomers is demonstrated to amplify the magnitude of circular dichroism.   
The realization of strong chirality in these oligomers demonstrates a new path to engineer optical chirality from planar devices using dark-field illumination.   
\end{abstract}

Dark-field (DF) imaging is one of the more straightforward avenues to directly observe resonances of single plasmonic nanoparticles, and has thereby found function in nanoparticle-assisted monitoring of few or single molecules\cite{McFarland2003, Ament2012, CarmenEstevez2014}  and spatially resolved monitoring of reactions \cite{Beuwer2015}.
Yet DF is less common than bright-field in the related pursuit of discriminating enantiomers of chiral molecules using more complex nanoparticle systems.  
The operation itself is portable to DF settings: existing investigations have sought to introduce a bias in the signed magnitude of local helicity density $h\!=\!\mathrm{Im}\{\mathbf{E}^*\!\cdot\mathbf{H}\}$, to promote a net difference in coupling strength between a nanoparticle antenna and oppositely handed chiral molecules\cite{tang2010optical, Yoo2015}.
The nanoparticle geometry is therefore a design freedom used to bias the helicity distribution generated under LCP (positive $h$) or RCP (negative $h$) illumination, for instance: it was shown that a nanoparticle antenna could foreseeably be designed to scatter only a single helicity of light\cite{Fernandez-Corbaton2016}.
However, the majority of experimental investigations have pursued an intermediary goal using true chiral geometries to produce {\it circular dichroism} (CD), a variation in the {\it extinction} (total power dissipated) from LCP and RCP illuminations, and then relying on some degree of helicity conservation in scattering\cite{Nieto-Vesperinas2015}
to create a helicity bias.   
Here we reveal that DF illumination is beneficial for this task because it: (i) enables simple {\sl planar chiral} nanoparticle arrangements to exhibit CD in a manner comparable to true chiral scatterers, and (ii) allows employment of desireable geometric symmetries that would otherwise be expected to suppress CD.

\begin{figure}[tbp!]
  \includegraphics[width=0.92\textwidth]{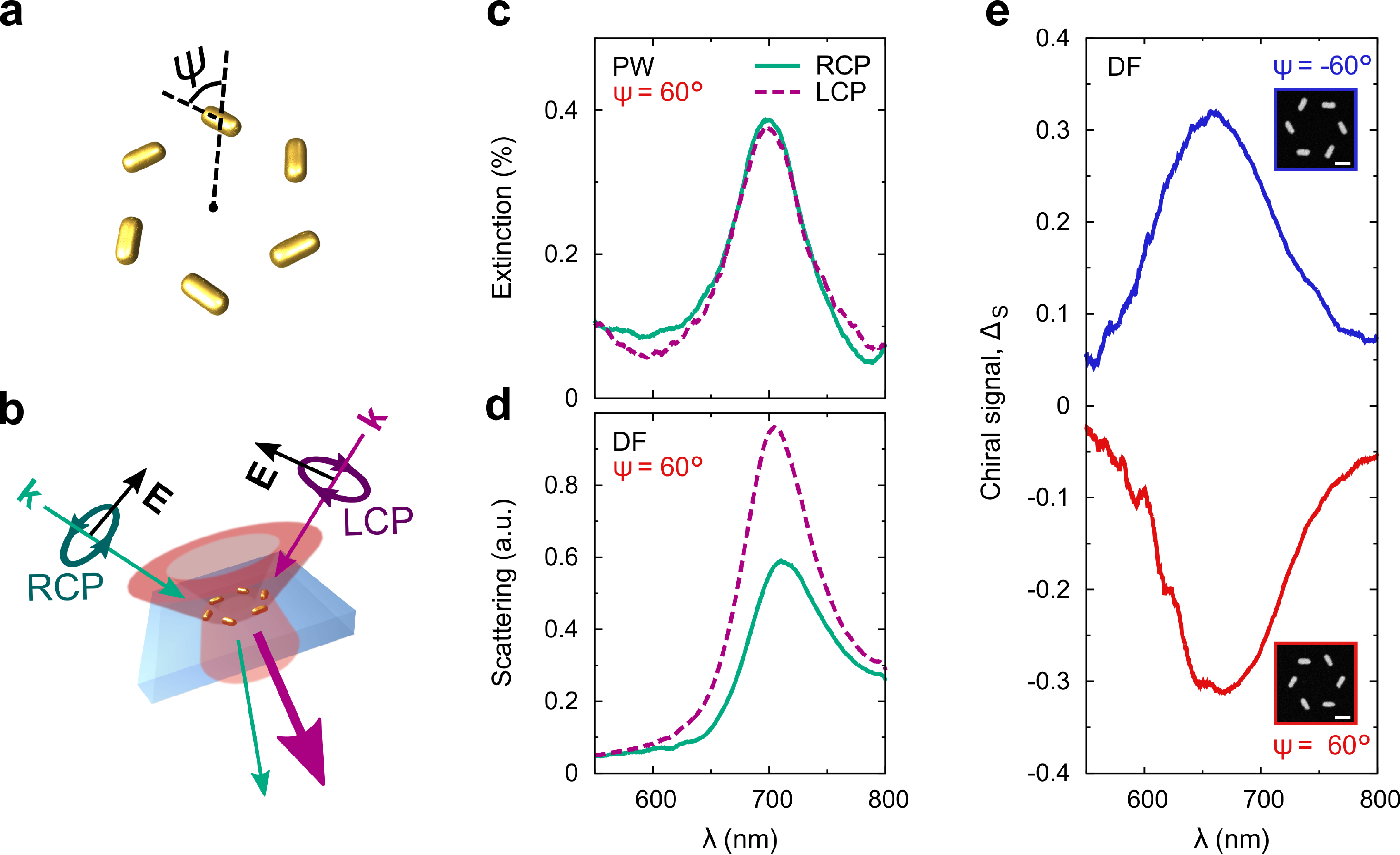}
  \caption{
  Schematics of (a) the planar gold nanorod oligomers and (b) the dark-field illumination.
  The experimentally measured (c) extinction spectra for plane waves (PW) and (d) scattering spectra for dark-field (DF) are shown for circularly polarized incident light with nanorod orientation set to $\psi\! =\! 60^{\circ}$.
  Results are also presented in terms of (e) chiral signal  \mbox{$\Delta_S\!=\!(S_{\scriptscriptstyle \mathrm{RCP}}-S_{\scriptscriptstyle \mathrm{LCP}})/(S_{\scriptscriptstyle \mathrm{RCP}}+S_{\scriptscriptstyle \mathrm{LCP}})$} for nanorod orientations \mbox{$\psi\!=\! -60^{\circ}$ and $60^{\circ}$} from DF illumination. 
  The dimensions of the nanorods were measured to be $35\times86$~nm, and the center-to-center distance between two opposing nanorods is $D\!=\!340$~nm.
  The scale bar in the insets is 100~nm.
  }
  \label{fig1}
\end{figure}
In Figure~\ref{fig1}, we show the experimental scattering spectra of a planar, gold nanorod oligomer with 6-fold rotational symmetry, depicted in (a), under right- and left-circular polarized (RCP and LCP) illuminations from (d) DF with a \mbox{0.8\! --\! 0.95~NA} condenser.
The extinction spectra of the same hexamer under RCP and LCP illuminations from normal incidence plane waves is also shown in Figure~\ref{fig1}(c) for comparison.
The DF scattering $S$ is measured as the transmission collected by a 0.6~NA objective lens, and this can be seen to vary significantly between RCP and LCP under DF illumination, whereas the extinction under normal incidence plane waves which was obtained by subtracting the transmission from unity does not vary.
The resonant peaks under DF are found at the wavelength of $704\,{\rm nm}$ and $713\,{\rm nm}$ in the scattering spectra of LCP and RCP of $\psi=60^{\circ}$, respectively. 
To characterize the chiral response, we define a chiral signal \mbox{$\Delta_S = (S_{\scriptscriptstyle \mathrm{RCP}}-S_{\scriptscriptstyle \mathrm{LCP}})/(S_{\scriptscriptstyle \mathrm{RCP}}+S_{\scriptscriptstyle \mathrm{LCP}})$}, as plotted in Figure~\ref{fig1}e.
Note that a single gold nanorod oligomers can exhibit strong chiral response over $\Delta_S=0.3$ {\sl while on resonance} at a wavelength of 670~nm.
In the Supplementary Material, further derivations show that the absence of chiral signal under normal plane wave illumination follows from having 3-fold or more rotational symmetry, encompassing substrate, once assuming each nanorod supports only a single dipole moment.
The scattering presented indeed appears to show significant CD, much like that which occurs for geometrically chiral antennas\cite{Decker2009, Kuzyk2012}, or with obliquely incident plane waves on achiral antennas that lack inversion symmetry\cite{Sersic2012}.
However, the gold nanorod oligomer in Figure~\ref{fig1} is planar and has inversion symmetry in the absence of substrate, meaning CD is largely not expected.  
Yet other forms of chiral signal, such as circular conversion dichroism, are also not expected due the presence of rotational symmetry, discussed further in the Supplementary material.   
The presence of at least 3-fold rotational symmetry is notably an example of desireable symmetry: it forbids scattering of oppositely handed fields in the transmission direction along its principle axis\cite{Fernandez-Corbaton2013}, thereby promoting the generation of a uniform helicity distribution under circularly polarized illumination.
The origin of the chiral signal we observe therefore warrants further attention, particularly given true three-dimensional chirality is challenging in fabrication, with subsequent interest in finding new avenues for chiral optical response in two-dimensional plasmonic structures\cite{Fedotov2006, eftekhari2012strong, du2015broadband, Hopkins2016,  ShuaiZu2016, khanikaev2016experimental}
In the coming discussion, we will show that true circular dichroism can be expected to occur under DF illumination due to a difference in the magnitude of the applied electric  projected onto the nanorods, which occurs due to ellipticity in the DF polarization when confined to plane of the nanorods.
A preferred orientation therefore exists when a nanobar's long axis is aligned with the major axis of the polarization ellipse, and {\sl the substrate} is shown to vary the angle of this polarization axis between LCP or RCP DF illumination.
This thereby leads to circular dichroism that depends on the orientation of the nanorods and {\sl only exists under DF illumination}.

\begin{figure}[htbp!]
\centering\includegraphics[width=0.6\textwidth]{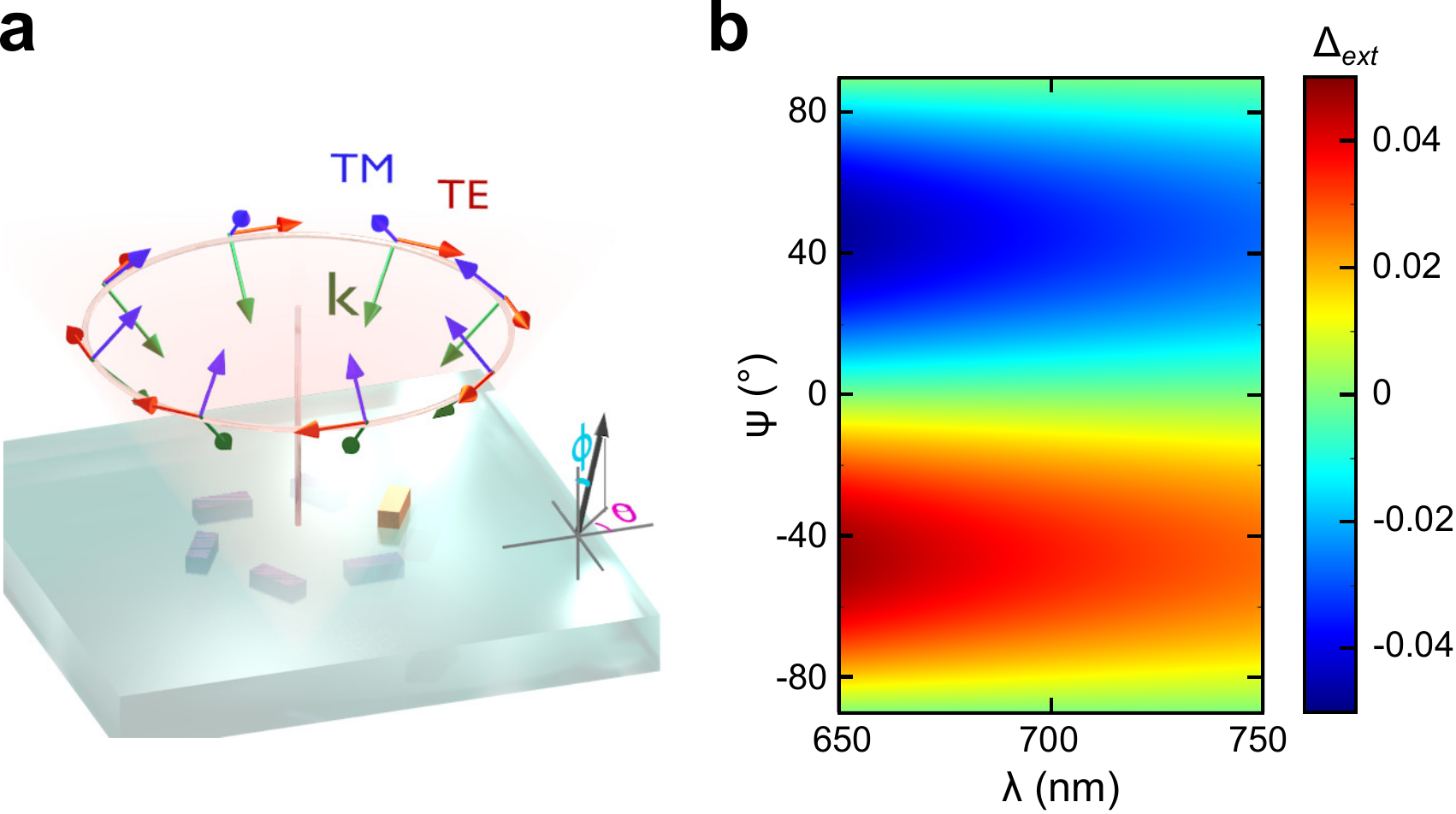}
\caption{ (a) Illustration of the annular ring of plane waves used in the dipole model Eq.~\ref{eq:TE} and \ref{eq:TM}, showing the TE and TM components of electric field polarization.
In (b) we show the predicted circular dichroism for extinction $\sigma$, expressed as a chiral signal \mbox{$\Delta_{ext}= (\sigma_{\scriptscriptstyle \mathrm{RCP}}-\sigma_{\scriptscriptstyle \mathrm{LCP}})/(\sigma_{\scriptscriptstyle \mathrm{RCP}}+\sigma_{\scriptscriptstyle \mathrm{LCP}})$} from the angular distributions of plane waves in (a), modelling a dark-field illumination.  
This was calculated using Eq.~\ref{eq:TE} and \ref{eq:TM}, with $\delta_x\!=\!170$~nm and $\delta_z\!=\!15$~nm.
}
\label{fig:theory}
\end{figure}

Let us consider a simplified model where we assume each gold nanorod behaves as an anisotropic point dipole aligned to the long-axis of the nanorod, and denote this orientation with a unit vector $\boldsymbol{\hat \psi}$.
We can recognize that the dipole moment amplitude of each nanorod is equal up to a phase shift with that of every other nanorod, because the propagation axis of the circularly polarized DF illumination is parallel the rotational symmetry axis of the oligomer.  
This means a symmetric rotation of the global coordinate system is equivalent to a phase shift of the applied field.
So we only need to consider a single nanorod, which we define as lying on the $\mathbf{\hat x}$-axis and shifted by a distance $\delta_x$.  
The substrate is placed a distance $\delta_z$  along $-\mathbf{\hat z}$, {\it i.e.} below the dipole.
To construct a DF illumination, we first consider the electric field from an annular ring of circularly polarized plane waves with incoming propagation ($\mathbf k$) vectors forming a circle in the tranverse $\mathbf k$-space, as illustrated in Figure~\ref{fig:theory}(a) for a polar incident angle $\phi$.
Our dipole will experience the projection of the applied electric field onto its orientation $\boldsymbol{\hat \psi}$, and this can be decomposed into the contributions from azimuthal, or transverse electric \mbox{$\mathbf E_{\scriptscriptstyle  \mathrm{TE}} \cdot \boldsymbol{\hat \theta}\cdot \boldsymbol{\hat \psi}\equiv \mathrm E_{\scriptscriptstyle  \mathrm{TE}} $}, polarizations and from polar, transverse magnetic \mbox{$\mathbf E_{\scriptscriptstyle  \mathrm{TM}} \cdot \boldsymbol{\hat \phi}\cdot \boldsymbol{\hat \psi}\equiv\mathrm E_{\scriptscriptstyle  \mathrm{TM}} $}, polarizations.
The TM field also includes an extra $\pm \frac{\pi}{2}$ phase shift to impose a circular polarization.
The resulting projection of electric field onto our dipole, will therefore given by: \mbox{$\mathrm {E}^{\scriptscriptstyle(\pm)} = \mathrm E^{\scriptscriptstyle(\pm)}_{\scriptscriptstyle  \mathrm{TE}} +\mathrm E^{\scriptscriptstyle(\pm)}_{\scriptscriptstyle  \mathrm{TM}}$}, with $\pm$ indicating LCP and RCP, respectively.
We provide derivation of $\mathrm E^{\scriptscriptstyle(\pm)}_{\scriptscriptstyle  \mathrm{TE}}$ and $\mathrm E^{\scriptscriptstyle(\pm)}_{\scriptscriptstyle  \mathrm{TM}}$ in the Supplementary Material, and state the result here. 
\begin{align}
\allowdisplaybreaks[1]
\mathrm E^{\scriptscriptstyle(\pm)}_{\scriptscriptstyle  \mathrm{TE}}& \;=\;   
\pm i \pi \big(J_0(-k \delta_x  \sin \phi )e^{\mp i \psi}+ J_2(-k \delta_x  \sin \phi )e^{\pm i\psi}\big)
\cdot (1 + {r}_{\scriptscriptstyle \mathrm{TE}} \,e^{2ik \delta_ z \cos \phi }) 
 \label{eq:TE}  \\[2ex]
\mathrm E^{\scriptscriptstyle(\pm)}_{\scriptscriptstyle \mathrm{TM}}& \;= \;
\pm i \pi \big(J_0(-k \delta_x  \sin \phi )e^{\mp i \psi}- J_2(-k \delta_x  \sin \phi )e^{\pm i\psi}\big)
\cos \phi
\cdot(1 - {r}_{\scriptscriptstyle \mathrm{TM}} \,e^{2ik \delta_ z\cos \phi }) 
 \label{eq:TM}
\end{align}
Here $ {r}_{\scriptscriptstyle \mathrm{TE}}$ and  $ {r}_{\scriptscriptstyle \mathrm{TM}}$ are Fresnel reflection coefficients for the air-substrate interface and $J_n$ is the $n^\mathrm{th}$ order Bessel function of the first kind.
A chiral signal can now be defined for the magnitude of the applied electric field projected onto $\boldsymbol{\hat \psi}$, being: \mbox{$\Delta_{\mathrm E}=(|\mathrm E^{\scriptscriptstyle(-)}|^2\!-|\mathrm E^{\scriptscriptstyle(+)}|^2)/(|\mathrm E^{\scriptscriptstyle(-)}|^2+|\mathrm E^{\scriptscriptstyle(+)}|^2)$}.
When using our dipole model, $\Delta_{\mathrm E}$ is equal to the chiral signal for the extinction $\sigma$ of the whole oligomer \mbox{$\Delta_{ext}= (\sigma_{\scriptscriptstyle \mathrm{RCP}}-\sigma_{\scriptscriptstyle \mathrm{LCP}})/(\sigma_{\scriptscriptstyle \mathrm{RCP}}+\sigma_{\scriptscriptstyle \mathrm{LCP}})$}, see Supplementary Material.
Circular dichroism therefore exists only due to a difference in the magnitude of the electric field projected onto $\boldsymbol{\hat \psi}$.
However, when there is no substrate \mbox{$\mathrm E^{\scriptscriptstyle(\pm)} = -(\mathrm E^{\scriptscriptstyle(\mp)})^*$}, hence \mbox{$|\mathrm E^{\scriptscriptstyle(+)}|^2 = |\mathrm E^{\scriptscriptstyle(-)}|^2  $}, and $\Delta_{\mathrm E}$ is {\sl precisely zero}.
It is only {\sl when we account for the substrate} that $\Delta_{\mathrm E}$ becomes  nonzero for this DF illumination, recognizing that $\Delta_{\mathrm E}$ remains zero for normal plane waves.
In Figure~\ref{fig:theory}(b) we integrate $\mathrm E^{\scriptscriptstyle(\pm)}$ over $\phi$ assuming a uniform amplitude plane wave distributed on an annular solid angle section in $\mathbf k$-space (\mbox{$\mathrm d \Omega\! =\! \sin\!\phi \mathrm\;d \theta \mathrm d \phi$}) truncated to the DF condenser's \mbox{0.8\! --\! 0.95~NA} range.
To deconstruct what is happening, we first note that the electric field lying in the plane of the nanorods is not necessarily circularly polarized when displaced from the propagation axis.
The electric field incident on the dipole will therefore have an elliptical polarization, leading to increased extinction when its major axis aligns with $\psi$.
Said another way, we cannot separate $\psi$ dependence as a common unitary factor from the sum of  $\mathrm E^{\scriptscriptstyle(\pm)}_{\scriptscriptstyle  \mathrm{TE}}$ and $\mathrm E^{\scriptscriptstyle(\pm)}_{\scriptscriptstyle  \mathrm{TM}}$.  
We can also see that interchanging the sign of $\psi$ in Eq.~\ref{eq:TE} and \ref{eq:TM} is equivalent to interchanging   \mbox{$|\mathrm E^{\scriptscriptstyle(+)}|^2 \leftrightarrow |\mathrm E^{\scriptscriptstyle(-)}|^2 $}.
As such, circular dichroism implies the magnitude of electric field projected onto $\psi$ is not equal to that projected onto $-\psi$.  
Given $\psi$ dependence cannot be separated as a common unitary factor from the sum of $\mathrm E^{\scriptscriptstyle(\pm)}_{\scriptscriptstyle \mathrm{TE}}$ and $\mathrm E^{\scriptscriptstyle(\pm)}_{\scriptscriptstyle \mathrm{TM}}$, circular dichroism can be expected whenever the applied electric field's major polarization axis is aligned with any \mbox{$\psi \neq 0,$ 90\degree}, being angles for which \mbox{$e^{\pm i\psi}=-e^{\mp i \psi}$}.
Explicitly, the Bessel functions in (\ref{eq:TE}) and (\ref{eq:TM}) are real valued, and without a substrate  (\mbox{$r_{\mathrm{TE}}= r_{\mathrm{TM}}=0$}): $|\mathrm E^{\scriptscriptstyle(\pm)}_{\scriptscriptstyle \mathrm{TE}}|$ is maximized at  $\psi = 0$ , and $|\mathrm E^{\scriptscriptstyle(\pm)}_{\scriptscriptstyle \mathrm{TM}}|$ is maximized at $\psi = \frac{\pi}{2}$. 
Note also that these two maxima are $\frac{\pi}{2}$ out of phase.  
The major polarization axis,  without a substrate, is therefore aligned to either $\psi = 0$ or 90\degree, hence no circular dichroism.
However, when introducing a substrate, the phase acquired in reflection due to $\delta_z$ makes the phase of $\mathrm E^{\scriptscriptstyle(\pm)}_{\scriptscriptstyle \mathrm{TE}}$ and $\mathrm E^{\scriptscriptstyle(\pm)}_{\scriptscriptstyle \mathrm{TM}}$ depend on $\phi$.  
The in-phase components when integrating over $\phi$ are therefore no longer going to uniformly be at $\psi = 0$ or 90\degree, indicating the major polarization axis will be rotated, hence leading to circular dichroism.  
It is worth recognizing that this ellipticity in the periphery of a beam resembles that from the optical spin-Hall effect\cite{Bliokh2015}, such from the reflection of a Gaussian beam at an interface.
Moreover, a linear-polarized Gaussian incident on an interface will generate a symmetric {\sl cross-polarization} in the portion of the beam periphery perpendicular to the original polarization axis.\cite{Bliokh2006}  
For a circular-polarized Gaussian, as the sum of two linear-polarized Gaussians, this effect can be expected to rotate the major axis of the transverse elliptical polarization out of alignment with $\psi=0, 90\degree$.
In the Supplementary Material, we also recalculate $\Delta_E$ from a DF modeled as a superposition of two Gaussians with NA~0.95 and NA~0.8, and show it predicts a circular dichroism magnitude closer to that seen in experiment.
We, therefore, conclude that the transverse orientation of applied electric fields, from a circularly polarized DF incident on a substrate, provides preferential coupling of LCP and RCP into the considered oligomers, associated with an angular dependence on the nanorod orientation $\psi$.

\begin{figure}[tbp!]
  \includegraphics[width=0.6\textwidth]{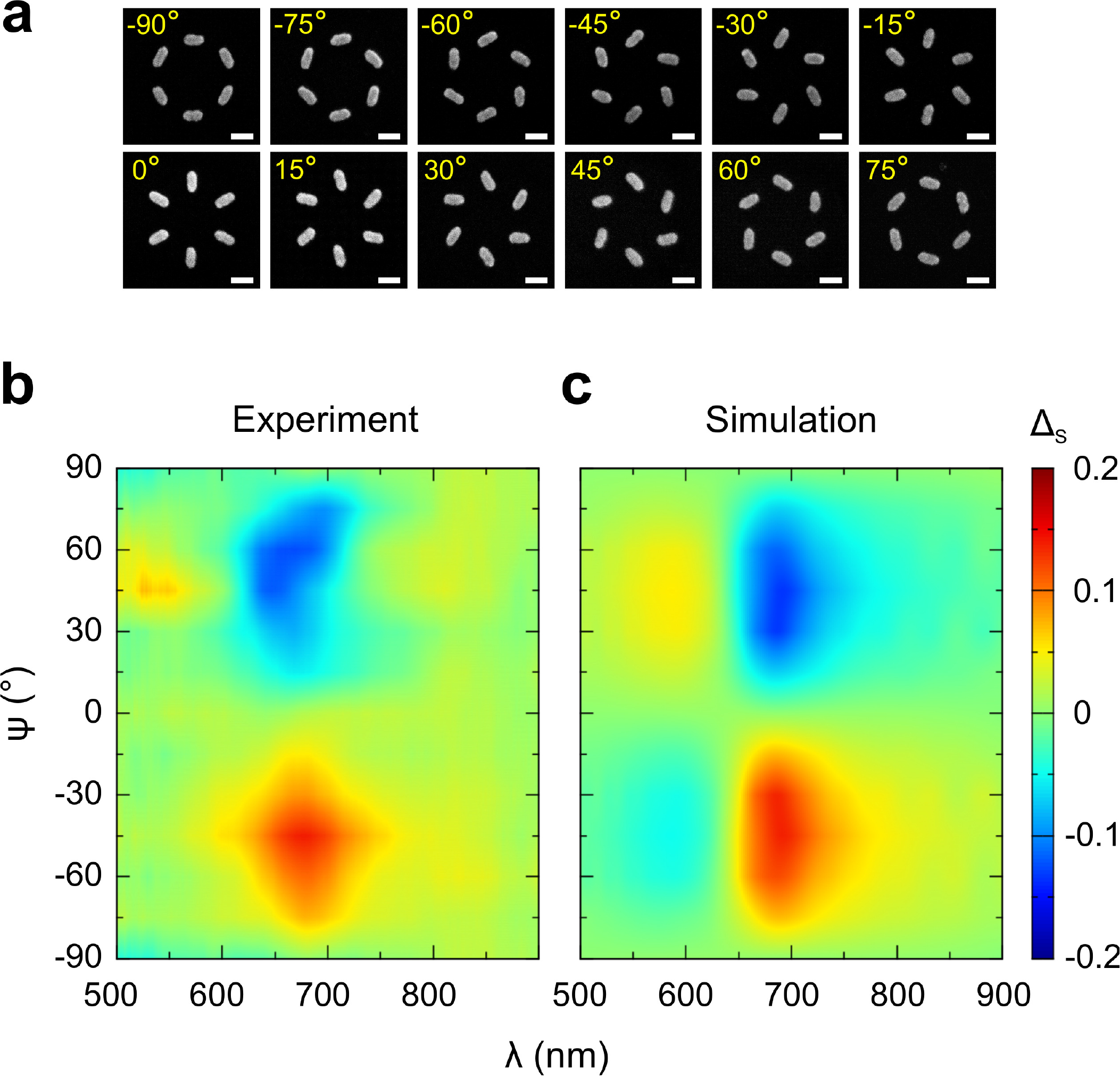}
  \caption{The variation of the observed chiral response of the plasmonic hexamers depending on the orientation angle $\psi$ of the gold nanorods.
  (a) SEM images of the plasmonic hexamers with different $\psi$, the scale bar is 100 nm. 
 (b)  Experimental measurements and (c) numerical simulations of the chiral signal $\Delta_S$ as functions of nanorod orientation $\psi$.
  }
  \label{figExpAngle}
\end{figure}
\begin{figure*}[tbp!]
    \includegraphics[width=0.92\textwidth]{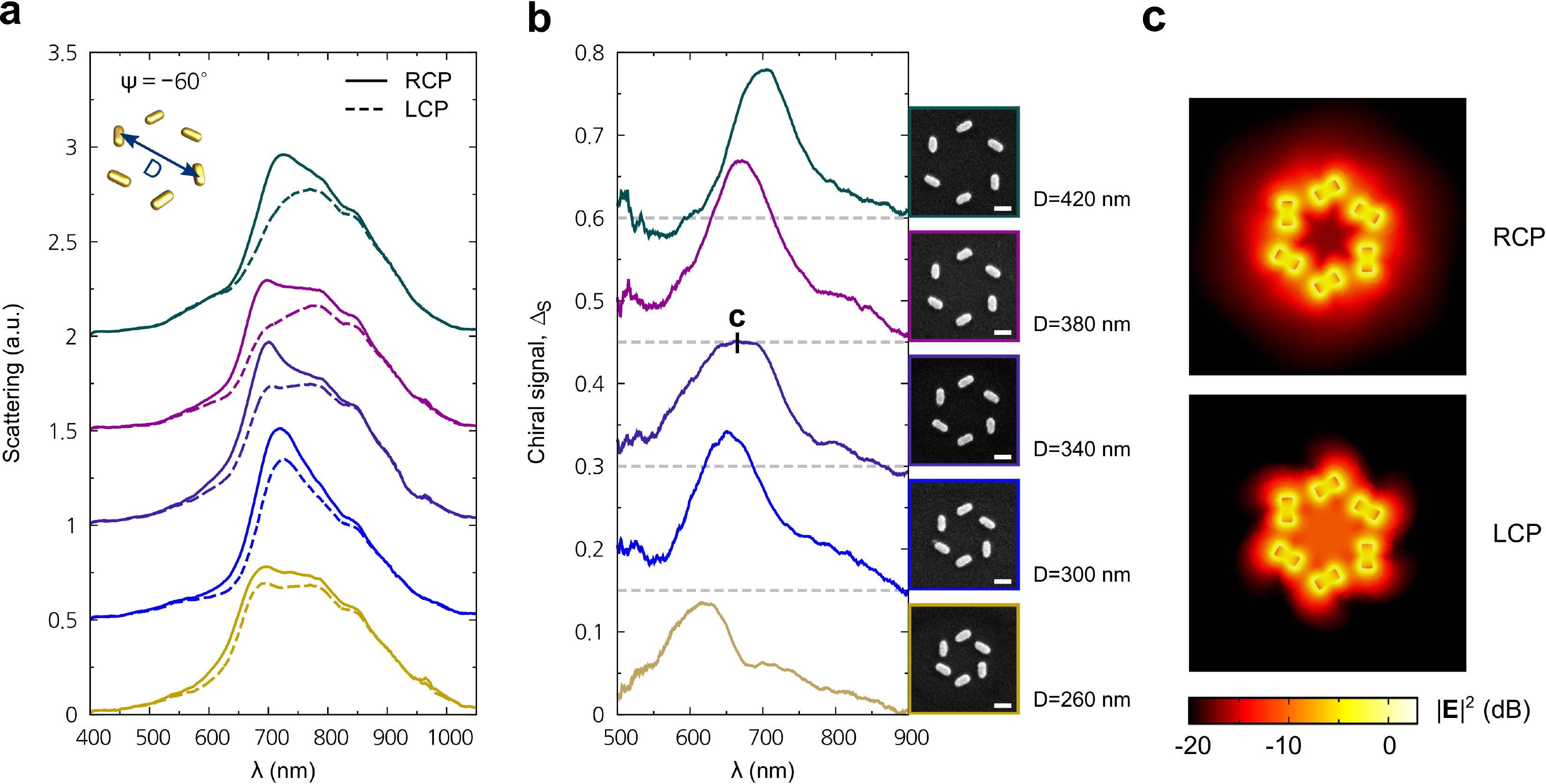}
    \caption{The measured (a) scattering and (b) chiral signal  spectra for different separations $D$ between opposing nanorods. 
    The SEM images have a scale bar of $100\,{\rm nm}$.
    Shown in (c) are simulated near-field profiles of $\left|{\rm \bf E}\right|^2$ in dB under RCP and LCP illumination, for the case where
    $D=340\,{\rm nm}$ and $\lambda=660\,{\rm nm}$.
    } \label{fig3}
\end{figure*}

We now show this angular dependence experimentally.
Gold nanorod oligomers were fabricated for the full $180^\circ$ range of possible $\psi$ in $15^\circ$ steps, as shown in Figure~\ref{figExpAngle}a. 
The dimensions of the fabricated nanorods were measured to be $43\times94$~nm.
The measured chiral signal $\Delta_S$ from experiment and from simulation using CST Microwave Studio, produce the maps presented in Figure~\ref{figExpAngle}b and c, respectively.
Both show the $\psi$-dependent circular dichroism that we expected from the dipole model, with no chiral response is observed at $\psi=0,90\degree$, and opposite sign of chiral signal observed for oppositely handed chiral oligomers.
The peak chiral signal here is also occurring in the vicinity of the oligomer's resonant frequency, which is expected because this is necessary to impose large anisotropy of the nanobars.
The measured chiral signal is thereby associated with significant differential scattering signal, rather than suppressing the denominator of $\Delta_S$, which has allowed measuring  strong chiral signal from even a {\sl single} oligomer.  
The difference in scattering magnitude  of several oligomers is shown later in Figure~\ref{fig3}a, and in simulation we also see a qualitative change in the local scattered electric field intensity depicted in Figure~\ref{fig3}c.
There is now a second degree of freedom we have over the chiral signal: the transverse displacement of nanorods $\delta_x$.
Moreover, referring to our model for Figure~\ref{fig:theory}, specfically Eq.~\ref{eq:TE} and \ref{eq:TM}, we assume that the nanorods are optically thin $\delta_z\approx0$, and then the wavelength dependence of the applied field projection is entirely due to the size of $\delta_x$ relative to the wavelength ($k\delta_x$).
We should therefore expect the transverse displacement of nanorods to shift the peak chiral signal, without having to change the nanorod dimensions.  
In Figure~\ref{fig3}, we present experimental investigation of nanorod oligomers when varying the center-to-center distance between opposing nanorods $D =2 \delta_x$ from $260\,{\rm nm}$ to $420\,{\rm nm}$ in increments of $40\,{\rm nm}$.
The corresponding scattering and chirality spectra are provided in Figure~\ref{fig3}a and b, respectively.
We do indeed observe a red-shift with increasing $D$ in the peak chiral signal from $\lambda=610\,{\rm nm}$ to $700\,{\rm nm}$, while the resonance wavelength of scattering signal remains largely stationary.
This thereby provides an effective way to engineer the wavelength of chiral DF scattering from a subwavelength planar plasmonic structure, without changing the physical dimension of the consisting metal nanorods, such as if constrained by fabrication or assembly procedure.

As mentioned, the resonant frequency of scattering in Figure~\ref{fig3} is also relatively constant, despite the changes to the oligomer dimensions, which suggests that we are observing a resonance that is dominantly determined by that of a single nanorod.
Additionally, the presented oligomers have a subwavelength physical footprint, owing to existence of quasistatic surface plasmon resonances when confined in highly subwavelength photonic devices\cite{barnes2003surface,kwon2010subwavelength,hwang2011plasmonic,hwang2013metal}.
These both suggest that we should be able to directly translate the chiral operation of a single oligomer into a homogenized metasurface platform.
So let us explore arrays of oligomers, as to show relevance of these concepts also toward pursuit of plasmonic metasurfaces for polarization control, and other applications for flat optics\cite{yu2014flat,zheng2015metasurface,YHwang2016APL,ee2016tunable} that capitalize on 2-dimensional fabrication.
\begin{figure}[tbp!]
  \includegraphics[width=0.6\textwidth]{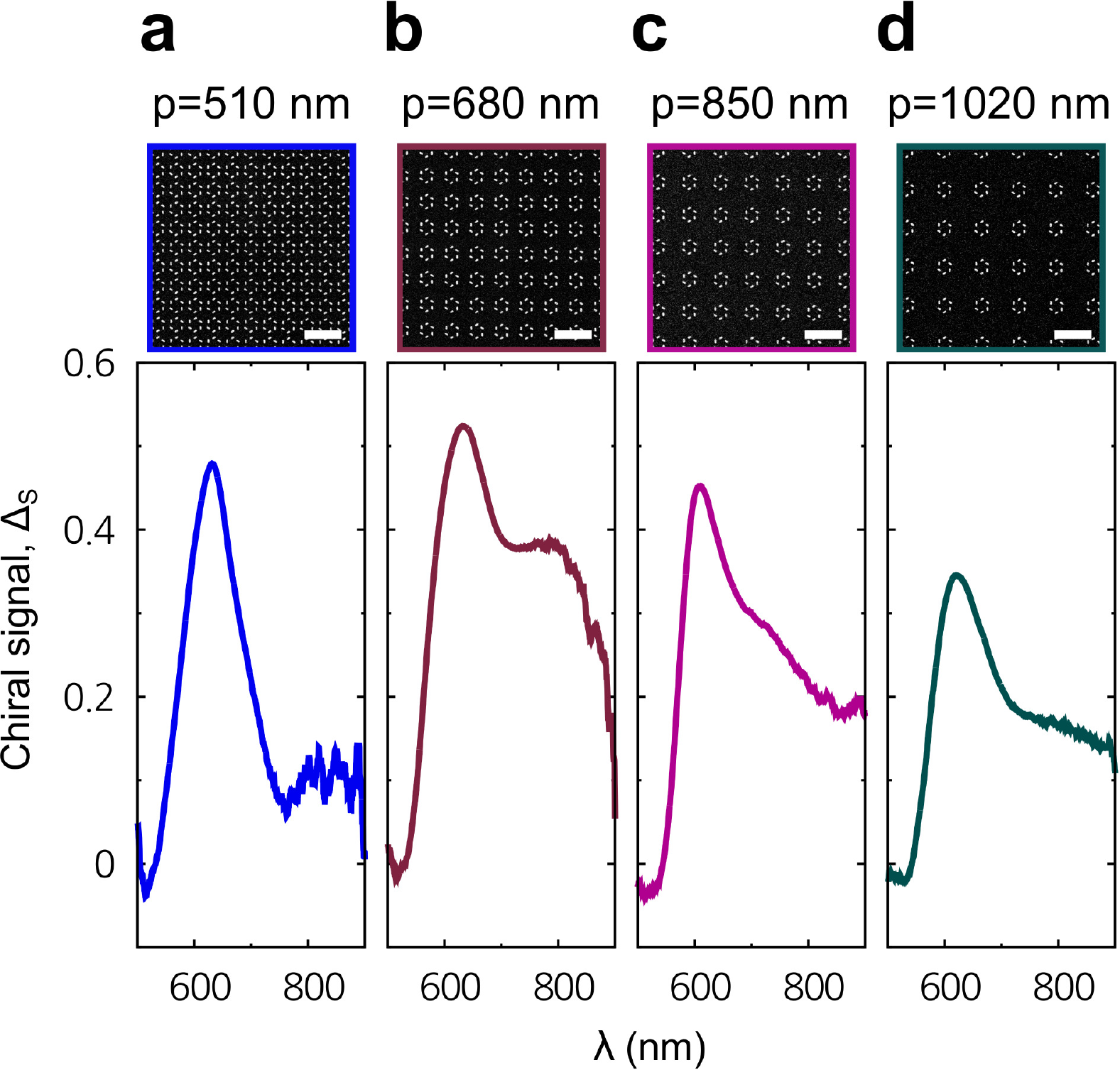}
  \caption{Chiral signal spectra of the arrays with the periods of (a) $510\,{\rm nm}$, (b) $640\,{\rm nm}$, (c) $850\,{\rm nm}$, and (d) $1020\,{\rm nm}$. $\psi$ and $D$ are fixed to be $-60^{\circ}$ and $340$~nm, respectively. 
  The scale bar in the SEM images is $1\,{\rm \mu m}$.
  }
  \label{fig4}
\end{figure}
In Figure~\ref{fig4}, we present experimental measurements of arrays of nanorod oligomers with lattice periods ranging from 510~nm to 1020~nm.
The strongest chiral response is observed for the array with  $p=680\,{\rm nm}$ whose maximum chiral signal $\Delta_S$ is over 0.5, which is approximately three times greater than the equivalent single oligomer in Figure~\ref{figExpAngle}.
Yet, simultaneously, no substantial wavelength shift is observed in the chiral signal peak, suggesting that the circular dichroism remains as the same field projection mechansim that occurs for the single oligomer.
As such, the lattice period is a parameter with which we can optimize the chiral signal magnitude without impacting wavelength shifts, and the nanoparticle spacing within a constituent oligomer is a parameter than governs wavelength shifts of the chiral signal.

To conclude, we have realized a subwavelength scale planar device that exhibits strong optical chirality using arrangements of plasmonic nanoparticles illuminated by dark-field. 
Theoretical analysis using a dipole model has explained the origin of the chiral response is a difference in the geometric projection of a nanorod onto the handed orientation of electric fields created by a circularly polarized dark-field incident on a glass substrate. 
The dependence of the orientation angle of the nanorods has been systematically demonstrated with experiments and numerical simulations.
It was also found that the wavelength range of the chiral response can be tuned by changing the distance between nanorods, and the magnitude can be amplified through the extension to closely packed, planar arrays of oligomers.   
Planar plasmonic chiral devices based on the demonstrated structures can therefore be employed in the place of true chiral geometries while also benefiting from simultaneous access to desireable geometric symmetries, including rotational symmetry considered here.
These results foreseeably benefit pursuits that require preferential generation of left- and right-handed helicity fields, such as nanoparticle-assisted discrimination of enantiomers.

\section{Methods}

 \noindent {\it \quad Nanofabrication}.\;
The plasmonic chiral oligomers were fabricated using 100 nm thick poly(methyl methacrylate) (PMMA) and 100 nm thick copolymer spin coated on a 500 $\mu$m thick glass substrate, then baked at 180 °C for 90 seconds.
The designed patterns were written on the electron beam resist using 100kV electron beam (Vistec EBPG5000plusES) and developed in methyl isobutyl ketone (MIBK):isopropyl alcohol (IPA) solution with the ratio of 1:3. Thereafter, electron beam evaporation was performed to deposit 2 nm thick germanium adhesion layer and 30 nm gold layer followed by a lift-off process using acetone. 
The nanorods were designed to be 90 nm long, 40 nm wide and 30 nm thick. 

 \noindent {\it \quad Optical measurement}. \;
Scattering spectra were measured under dark-field illumination on the plasmonic structures to examine their chirality. 
Broadband light from a halogen lamp was right- or left-circularly polarized (RCP or LCP) by a linear polarizer and an aligned quarter wave plate.
The RCP and LCP light were focused by a dark-field condenser with the range of the numerical aperture of 0.80-0.95 which means the light in the corresponding solid angle was exclusively illuminated on the plasmonic oligomers as shown in Figure~\ref{fig1}a.
The forward scattered light from the oligomers was collected by an object with the numerical aperture of 0.60 in order to avoid the direct transmission.
The collected light was resolved for the wavelength by a monochromator and then imaged on a cooled CCD (Princeton Instrument PIXIS).

 \noindent {\it \quad Numerical simulation}.\;
Simulations were performed using (Figures~\ref{fig3} and~\ref{fig4}) CST Microwave Studio for Finite Difference Time Domain (FDTD), and  (Supporting Information) Comsol Multiphysics for Frequency Domain Finite Element Method; all methods were found to provide comparable results.  
The dark-field illumination was modeled as a superposition of two paraxial Gaussian beams with NA 0.8 and 0.95; the respective electric field amplitudes were balanced to remove the electric field from the propagation axis in the far-field.
The signal collected by a 0.6~NA objective lens in transmission was modeled as a solid angle integral of the radially outward component of the Poynting vector for the {\sl scattered} field, evaluated in glass substrate at distance from the oligomer origin equal to the largest free-space wavelength.
Scattered field in CST was defined as the field radiated when imposing a volumetric box of dark field illumination enclosing only the oligomer.
Scattered field in Comsol was defined by first simulating the resulting field distribution from the dark field incident on the substrate interface without nanoparticles, then imposing the resulting field internal to the previously absent nanoparticles as an external field in a second simulation to calculate the scattered field. 
Permittivity of the glass substrate was taken to be 2.3, the permittivity of gold was taken from Johnson and Christy\cite{Johnson1972}, and the size of the nanorod was set at $90\times40\times30$~nm.

\section{Acknowledgements}
The work is supported by the leading talents of Guangdong province program No. 00201505 and the Natural Science Foundation of Guangdong Province under No. 2016A030312010. 
This work was performed in part at the Melbourne Centre for Nanofabrication (MCN) in the Victorian Node of the Australian National Fabrication Facility (ANFF).
BH acknowledges the discussions on theory and simulation with A.E. Miroshnichenko, Z. Fan, S. Trendafilov and M.R. Shcherbakov, the advice of F. Demming and C. Kremers  (CST) on simulations using CST Microwave Studio, and the support from A.E. Miroshnichenko, Y.S. Kivshar and G. Shvets.

\begin{suppinfo}
Further experimental results are presented showing the dependence of the nanorod oligomers on the number of constituent nanorods.
For the considered class of nanorod oligomers, we provide analysis on the consequences of their symmetry, analysis of the collective eigenmodes in a dipole model, and derivation of Eq.~\ref{eq:TE} and \ref{eq:TM}.  
\end{suppinfo}

\bibliography{chiral}

\end{document}


\newpage

\section{Dipole excitation from annular illumination}
Here we will derive the expressions presented in Equations (1) and (2) the main text, which describe the electric field from an annular ring of circularly polarized plane waves projected onto a point dipole oriented at angle $\psi$.
Transverse Electric (TE) and Transverse Magnetic (TM) polarization components are defined as shown in Figure~2a of the main text. 
\begin{align}
\allowdisplaybreaks[1]
\mathrm E^{\scriptscriptstyle(\pm)}_{\scriptscriptstyle  \mathrm{TE}}& \;=\;   \int \limits_0 ^{2\pi}\!\!
\scriptbelow{\underbrace{\sin(\psi-\theta)}}
{\begin{array}{c}\text{\tiny nanorod alignment}\\[3pt] \text{\tiny with TE}\end{array}}
\!\!\!\!\cdot {e^{i k_x \Delta}\cdot \!\!\!\!\!\! \scriptbelow{ \underbrace{e^{\pm i \theta}}}
{\begin{array}{c}\text{\tiny rotating}\\[3pt] \text{\tiny polarization}\end{array}}
}\!\!\!\!\!
\cdot \scriptbelow{\underbrace{ (1 + \mathrm{r}_{\scriptscriptstyle {TE}} \,e^{2ik_z\delta}) }}
{\begin{array}{c}\text{\tiny incident and}\\[3pt] \text{\tiny substrate reflection}\end{array}} \; \mathrm{d}\theta
 \label{eq:TE}  \\[2ex]
\mathrm E^{\scriptscriptstyle(\pm)}_{\scriptscriptstyle \mathrm{TM}}& \;= \;\int \limits_0 ^{2\pi}\! \pm i \,
\scriptbelow{\underbrace{\cos(\psi - \theta)\cos(\phi)}}
{ \begin{array}{c}\text{\tiny nanorod alignment}\\[3pt] \text{\tiny with TM}\end{array}}
\cdot {e^{i k_x \Delta} \cdot \!\!\!\!\!\! \scriptbelow{ \underbrace{e^{\pm i \theta}}}
{\begin{array}{c}\text{\tiny rotating}\\[3pt] \text{\tiny polarization}\end{array}}
}\!\!\!\!\!
\cdot \; \scriptbelow{\underbrace{ (1 - \mathrm{r}_{\scriptscriptstyle {TM}} \,e^{2ik_z\delta}) }}
{\begin{array}{c}\text{\tiny incident and}\\[3pt] \text{\tiny substrate reflection}\end{array}}\; \mathrm{d}\theta
 \label{eq:TM}
\end{align}
Here $k_x = k\cos(\phi)\cos(\theta)$, $k_z = k\sin(\phi)$, and $\mathrm{r}_{\scriptscriptstyle TE}$ and $\mathrm{r}_{\scriptscriptstyle TM}$ are the standard Fresnel reflection coefficients for an interface between two dielectric media with refractive indices $n_1$ and $n_2$.
\begin{align}
\mathrm{r}_{\scriptscriptstyle TE} &= \frac{n_1 \cos\phi-n_2\cos\phi'}{n_1\cos\phi+n_2\cos\phi'}, \quad \mathrm{r}_{\scriptscriptstyle TM} = \frac{n_2 \cos\phi-n_1\cos\phi'}{n_1\cos\phi'+n_2\cos\phi}
\end{align}
With  $n_1 \sin \phi = n_2 \sin \phi'$.
Note that modeling the reflection interaction using Fresnel coefficients is an approximation to treat the dipole moment of the nanorod on substrate as a point dipole floating above the surface.
The $e^(\pm i theta)$ phase factor is to appropriately rotate the polarizations over the whole loop to be simultaneously parallel, utilizing the fact that phase shifts are equivalent to transverse rotations of the polarization of circularly polarized plane wave.
The integral over $\theta = [0,2\pi)$ prescribes excitation from a ring of the TE and TM plane waves in $\mathbf{k}$-space.
This integral can be evaluated analytically in terms of Bessel functions, leading to the expressions presented in the main text.
\begin{align}
\allowdisplaybreaks[1]
\mathrm E^{\scriptscriptstyle(\pm)}_{\scriptscriptstyle  \mathrm{TE}}& \;=\;   
\pm i \pi \big(J_0(-k \delta_x  \sin \phi )e^{\mp i \psi}+ J_2(-k \delta_x  \sin \phi )e^{\pm i\psi}\big)
\cdot (1 + {r}_{\scriptscriptstyle \mathrm{TE}} \,e^{2ik \delta_ z \cos \phi }) 
 \label{eq:TE2}  \\[2ex]
\mathrm E^{\scriptscriptstyle(\pm)}_{\scriptscriptstyle \mathrm{TM}}& \;= \;
\pm i \pi \big(J_0(-k \delta_x  \sin \phi )e^{\mp i \psi}- J_2(-k \delta_x  \sin \phi )e^{\pm i\psi}\big)
\cos \phi
\cdot(1 - {r}_{\scriptscriptstyle \mathrm{TM}} \,e^{2ik \delta_ z\cos \phi }) 
 \label{eq:TM2}
\end{align}
where $J_n$ is the $n^\mathrm{th}$ order Bessel function of the first kind.

\section{Comparison to Gaussian dark-field illumination}
\begin{figure}[!ht]
\includegraphics[width=0.92\textwidth]{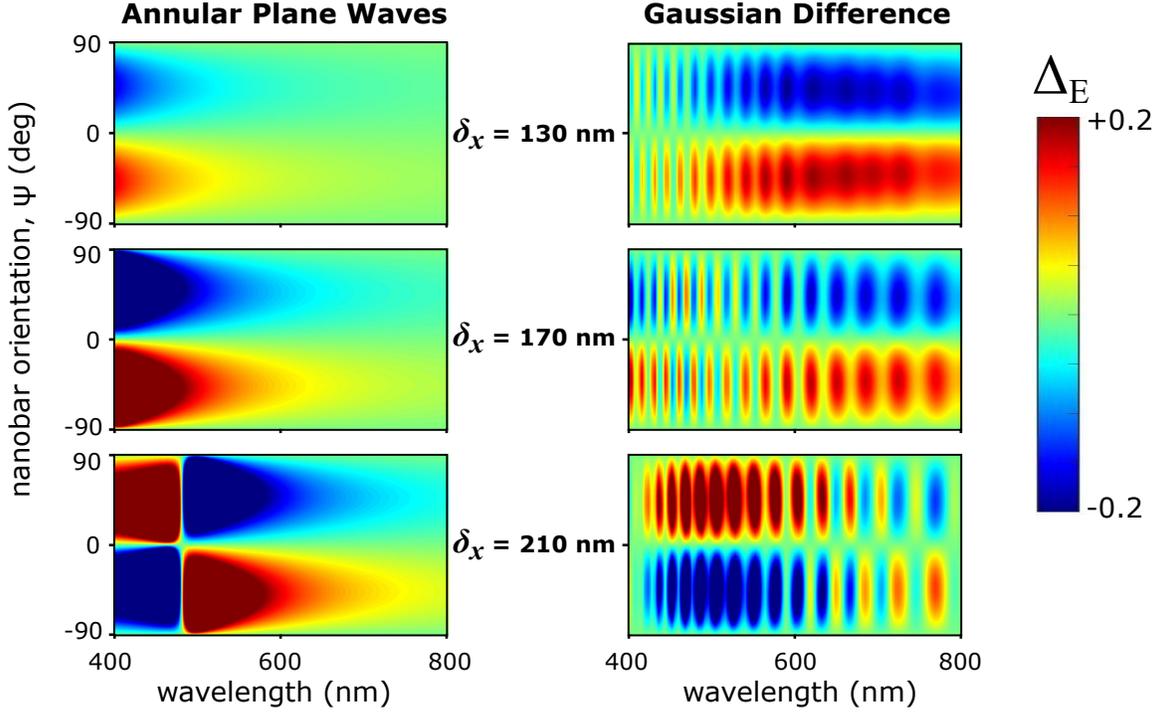}
\caption{Calculations of the chiral signal in excitation magnitude $\Delta_E$ experienced by an anisotropic dipole  oriented at an angle $\psi$ under dark-field illumations on a permittivity 2.3 substrate at $\delta_z=15~\mathrm{nm}$ from the dipole.  
The two models for dark-field are (left) uniform distributions of plane waves between NA 0.8 and 0.95, and (right) the difference of two circularly polarized Gaussian beams with NA 0.8 and 0.95. 
The transverse displacement if the dipole from the dark-field beam axis is varied from 130~nm to 210~nm, as to match Figure~4 of the main text. }
\label{fig:DFcomparison}
\end{figure}

The previous section described a simplified dark-field illumination from a uniform distribution of plane waves, and this was used in Figure~2b of the main text, which evaluated a solid angle of uniform illumination for NA from 0.8 and 0.95.  
Conversely, the fullwave numerics presented in the main text Figures~4 and 5 describe the dark-field illumination as a difference of Gaussians with NA 0.8 and with NA 0.95.  
Here we show a comparison between the excitation that a point dipole, orientated with a transverse angle $\psi$, experiences under both descriptions of dark-field illumination.
 In Figure~\ref{fig:DFcomparison}, we plot a chiral signal from the intensity of the electric field interacting with the oriented dipole, \mbox{$\Delta_{\mathrm E}=(|\mathrm E^{\scriptscriptstyle(-)}|^2\!-|\mathrm E^{\scriptscriptstyle(+)}|^2)/(|\mathrm E^{\scriptscriptstyle(-)}|^2+|\mathrm E^{\scriptscriptstyle(+}|^2)$}, while varying the transverse distance $\delta_z$.
Note that the chiral signal presented in Figure~\ref{fig:DFcomparison} is only representative of the chiral signal observed in experiment and full-wave numerics when the nanorod is resonant to validly neglect the dipole moment of each nanorod along the short axis, which is approximately a range of validity for 650-700~nm.
In this regime, it is clearly seen that the chiral signal expected under the Gaussian description of dark-field illumination is much greater than that when using the  uniform annular distribution of plane waves.

\section{Symmetry analysis and chiral optical scattering}
\label{sec:Symm}
Geometric chirality refers to objects whose mirror images are distinguishable because they cannot be superimposed using any combination of rotation or translation operations on their coordinate systems, which is equivalent to the absence of any symmetric planes of reflection, axes of improper rotation, or centers of inversion.
However, {\sl optical chirality} refers to the absence of these same symmetries in optical field distributions, including LCP and RCP plane waves.
In this regard, the symmetry of scattered electromagnetic fields will match that of the physical object superimposed with the externally applied fields; this follows given the overlap volume is able to determine all scattered fields using an integral equation method to solve for induced currents\cite{yaghjian1980electric}.  
One must therefore generally recognize that even achiral objects can be superimposed with applied fields such that their configuration is chiral, thereby allowing distinction of optical scattering from mirrored configurations.
Our oligomer, when considered in isolation to the glass substrate, is {\sl not} geometrically chiral due to a plane of mirror symmetry perpendicular to the principle rotation axis.
Our measurements therefore resemble that of chiral scattering phenomena from achiral scattering objects.  
To this extent, {\sl circular dichroism} and  {\sl circular conversion dichroism} are known chiral effects that have previously been observed in the scattering from achiral nanostructures, respectively corresponding to a dependence of co-polarized and cross-polarized transmission on LCP or RCP illumination.
\begin{itemize}
\item {\sl Circular dichroism}: a difference in the power lost by a plane wave depending on whether it has left- or right-circular polarization. 
\item {\sl Circular conversion dichroism}: a difference in the transmitted cross-polarization depending on whether the plane wave has left- or right-circular polarization.
\end{itemize}
Yet the considered oligomers have additional symmetries intended to suppress these two phenomena: (i) $n$-fold rotational symmetry with $n\geq 3$, and (ii) point inversion symmetry when $n$ is even. 
The specific consequences being:
\begin{enumerate}[label=(\roman*),align=Center, leftmargin=\parindent]
\setlength\itemsep{-1ex}
\item {\sl Discrete rotational symmetry}.  
Generation of cross-polarized transmission is forbidden in any circularly polarized plane wave propagating parallel to an $n$-fold discrete rotational symmetry axis when $n\geq3$\cite{Fernandez-Corbaton2013}. 
While a plane wave decomposition for a dark field does contain obliquely propagating components.
In the next section, we assume each nanorod behaves as an electric dipole with orientation fixed along its long-axis, and show that only a single doubly degenerate eigenmode that is excited by a LCP or RCP dark-field or plane wave illuminations.  
Circular conversion dichroism is therefore suppressed to the extent of cross polarization generation in the periphery of our collecting objective lens, and equally for both plane wave and dark field, given both couple to the same degenerate resonance.
\item {\sl Point inversion symmetry}.  
The scattered fields from a plane wave illumination are equivalent under the symmetric inversion of the coordinate system to that from the spatially inverted plane wave.    
Combined with reciprocity, which we can consider as the equal total losses experienced by illumination field distributions $\mathbf{E_0}$ and $\mathbf{E_0}^*$, LCP and RCP plane waves must experience equal losses along any propagation direction. 
In other words, we make the following relationship for circularly polarized plane waves:
\begin{align}
\mathbf{E_0}\equiv\Bigg( \!\!\begin{array}{c} 
\pm 1 \\  i
\end{array}\!\!\Bigg) e^{i \mathbf k \cdot \mathbf r}
\scriptbelow{\longleftrightarrow}{\begin{array}{c}\text{\scriptsize spatial} \\ \text{\scriptsize inversion}\end{array}}
\Bigg( \!\!\begin{array}{c} 
\mp 1 \\ - i
\end{array}\!\!\Bigg) e^{- i \mathbf k \cdot \mathbf r}
\scriptbelow{\longleftrightarrow}{\begin{array}{c}\text{\scriptsize \text{reciprocal}} \\ \text{\scriptsize plane waves}\end{array}}
\Bigg( \!\!\begin{array}{c} 
\mp 1 \\  i
\end{array}\!\!\Bigg) e^{i \mathbf k \cdot \mathbf r}
\label{eq:elephant}
\end{align}
\end{enumerate}
\noindent
As we observe no chiral signal from plane wave illumination, we can expect that circular dichroism is the cause of our chiral signal, despite (ii).   
More specifically, let us consider the impact of the substrate.  
A normally incident LCP or RCP plane wave reflected off the substrate is still circularly polarized, and circular dichroism therefore remains forbidden by to the same  inversion symmetry constraints of the oligomer in (ii).
However, an obliquely incident LCP or RCP plane wave is no longer circularly polarized after reflection, which creates an avenue to bypass the constraints of inversion symmetry of the oligomer.

\section{Number of nanorods}
\begin{figure}[h!]
  \includegraphics[width=0.6\textwidth]{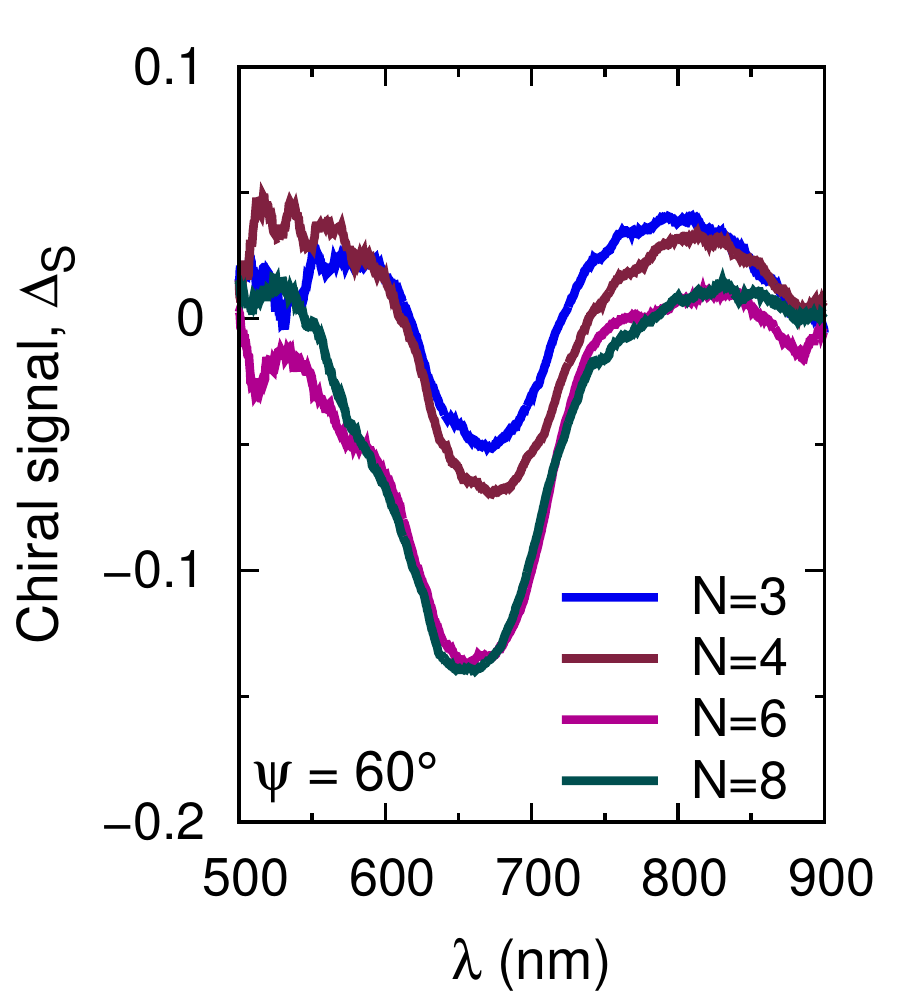}
  \caption{
  Measured chiral signal $\Delta_S$ spectra of the plamonic oligomers with different number of nanorods: $N=3, 4, 6,$ and $8$.
  $\theta = 60^{\circ}$ for all oligomers.
  }
  \label{figS2}
\end{figure}

In Section~\ref{sec:Symm}, it was argued that the presence or absence of inversion symmetry is not important to circular dichroism we are observing, despite the absence of inversion symmetry being necessary for oblique incidence circular dichroism on achiral structures, see (\ref{eq:elephant}).   
Here we support this argument experimentally, by showing that the considered gold nanorod oligomers exhibit qualitatively the same circular dichroism signal irrespective to the number of nanorods they contain.
Planar oligomers with 3, 4, 6 and 8 gold nanorods are shown in Figure~\ref{figS2}. 
The case of 3 nanorods does not have inversion symmetry, even if we neglect the presence of the substrate.  
Chiral response in the scattering from all these oligomers were experimentally measured under dark-field illumination, and the observed chiral signal spectra $\Delta_S$ are qualitatively similar.

\section{Eigenmodes of nanorod oligomers}

\subsubsection*{Definition of eigenmodes}
One can generally define eigenmodes from the induced current description of scattering.\cite{yaghjian1980electric}
\begin{align}
 - i \omega [ \boldsymbol{\bbar \epsilon} (\mathbf{r}) - \epsilon_0] \! \cdot  \mathbf{E}_{0}(\mathbf{r})   =&\,{\mathbf{J}(\mathbf{r})}  - \frac{k^2}{\epsilon_0} [  {{\epsilon}(\mathbf{r})  - \epsilon_0}]  \bigg( \! \text{\scriptsize P.V.} \! \! \int \!\!\Big [ \mathbf{\bbar{G}}(\mathbf{r},\mathbf{r'}) - \mathbf{\bbar{\,L}} \frac{\delta (\mathbf{r}-\mathbf{r'})}{k^2 } \Big] \!\cdot  \mathbf{J}(\mathbf{r'})\;{\mathrm{dr'}^3}\bigg)   \label{eq:current equation}
\end{align}
Here $\mathbf r$ is a position vector, $\epsilon$ is the permittivity distribution of the scattering object, $k$ is the wavenumber, $\mathbf{E_0}$ is the applied field distribution, and $\mathbf J$ is the induced current distribution.  
The $\text{\small P.V.}$ implies a principal value exclusion of the location $\mathbf{r'}=\mathbf{r}$ when performing the integration, and $ \mathbf{\bbar{\,L}}$ is the source dyadic necessary to account for the shape of this exclusion.  Finally $\mathbf{\bbar{G}}$ is the dyadic Green's function.
 \begin{align}
 {\mathbf{\bbar{G}}}(\mathbf{r},\mathbf{r'})
 &=    \left [\mathbf{\bbar{I}}+ \frac{1}{k^2} \boldsymbol{\nabla \nabla}\right] \frac{e^{i k R}}{4 \pi   R}  = \frac{ e^{i k R }}{4 \pi  R}\bigg[ \Big(1 + \frac{i}{k R}  - \frac{1}{k^2 R^2} \Big)\, \mathbf{\bbar I} - \Big(1 + \frac{3 i}{k R}  - \frac{3}{k^2 R^2}\Big) \mathbf{\hat n}\mathbf{\hat n}^\mathrm{T} \bigg]   \label{eq:G0_1} 
\end{align}
where $R = \left| \mathbf{r}-\mathbf{r'}\right|$ and $\mathbf{\hat n}$ is the unit vector pointing from $\mathbf{r'}$ to $\mathbf{r}$, in other words: $R \, \mathbf{\hat n}= \mathbf{r}-\mathbf{r'}$.
A single eigenmode  $\boldsymbol{ v}$ of this operator relating fields to induced currents, is a stable current oscillation \mbox{($\mathbf J = \boldsymbol v \rightarrow \mathbf{E_0} =\lambda \boldsymbol v$)}, with eigenvalue $\lambda$.
The inverse eigenvalue $\lambda^{-1}$ therefore represents a particular polarizability for the given system, and resonances occur at complex frequencies when an eigenvalue is equal to zero, denoting a nontrivial solution for $\mathbf{E_0}=0$.  

The scatterer we consider is a ring of $n\geq3$ anisotropic nanorods arranged with $n$-fold discrete rotational symmetry ($C_n$), and we will assume each nanorod only supports an electric dipole moment along its long axis. 
An eigenmode of the coupled dipole system is thereby a set of concatenated dipole moments $|p\rangle$ to the corresponding set of applied electric fields $|E\rangle$ at the location of each dipole.
In other words, (\ref{eq:current equation}) reduces to a matrix equation of the form:
\begin{align}
\mathbf{\mathbf{\bbar M}}\,|p\rangle =   |E\rangle\;. \label{eq:couplingM}
\end{align}
An eigenmode of coupling matrix $\mathbf{\bbar M}$ is a dipole moment profile $|v\rangle$ that is induced by a proportional electric field profile given by $\lambda |v\rangle$.

\subsubsection*{Use of $C_n$ symmetry}

Here we consider $n$-fold discrete rotational symmetry ($n\geq3$), and the set of symmetry operations (rotations by multiples of $\frac{2\pi}{n}$), which form the point symmetry group $C_n$.
We will use symmetry arguments to deduce the complete set of eigenmodes for our described oligomer geometries in the dipole model.
There will will be at most $n$ eigenmodes for an oligomer of $n$ nanorods with $C_n$ symmetry: $n$ is the maximum rank of the coupling matrix $\mathbf{\bbar M}$ in (\ref{eq:couplingM}), given it is the number of orthogonal dimensions in $|E\rangle$ and $|p\rangle$ when the orientation of each dipole is fixed.
So let us begin by defining some basis vectors for the dipole response (for $|p\rangle$) of our oligomer corresponding to each irreducible representation of $C_n$.
In this regard, there are only rotation symmetry operations and only one-dimensional irreducible representations in the symmetry group$C_n$.
This means that the basis vectors (for $|p\rangle$) corresponding to each irreducible representation, transform under symmetric rotations as scalar multiplication of basis vectors, denoted by the corresponding indices in character tables.  
For instance, see the character table for $C_3$ presented in Table~\ref{tab:C3}.
\begin{table}[!ht]
\centerline{
\begin{tabular}{|c|c|c|c||c|}
\hline & & & & \\[-2ex]
& \large$\hat E$ &\large $\hat C_3$ &\large $(\hat C_3)^2$ & {\small examples for} $[x,y,z]$\\
\hline & & & & \\[-2ex]
\large $A$ & 1 & 1& 1& $z$\\[1ex]
\large $E$ & $\bigg\{\begin{array}{c}1\\ 1\end{array}$ & $\begin{array}{c}\phi \\ \phi ^*\end{array}$ & $\begin{array}{c}\phi ^*\\ \phi \end{array}$ & $\begin{array}{c}x + i y\\ x - i y\end{array}$\\[1.5ex]\hline
\end{tabular}}
\caption{Character table for the $C_3$ symmetry group (3-fold rotational symmetry).  Rows denote irreducible representations, columns denote symmetry operations ($\hat E$ is the identity, $\hat C_3$ is a rotation by $\frac{2\pi}{3}$), indices are the character of the corresponding matrix representation for the symmetry operation ($\phi = e^{\frac{i 2 \pi}{3}}$)}.
\label{tab:C3}
\end{table}
Moreover, each irreducible representation of a given $C_n$ differs only by the phase acquired by rotation of the coordinate system by $\frac{2 \pi}{n}$ ($n\geq3$).
The only constraint is that a basis vector must acquire a multiple of $2\pi$ phase after $2\pi$ of geometric rotation in $\frac{2\pi}{n}$ increments.  
This constrains the dipole moment profiles of basis vectors to either:
\begin{itemize}[leftmargin=\parindent ]
\item{0 phase acquired under a $\frac{2\pi}{n}$ geometric rotation, which can occur if all dipoles are in phase, or if each dipole is $\pi$ out of phase with its neighbours ($n$ must be even).}
\item{$\pm \frac{2 \pi m}{n}$ phase acquired under a $\frac{2\pi}{n}$ geometric rotation, where $m=1,2,...\,\mathcal{M}$, and $\mathcal M$ is the largest positive integer $\frac{2\pi M}{n}< \pi$.\footnote{More formally, the maximum $M$  exists because any phase profile for $m' > M$ is equivalent to a profile of $m \leq \mathcal M$: we can define $m' = \mathcal{M}  + m$ for which $e^{\frac{\pm 2\pi (\mathcal M + m)}{n}} =e^{\frac{\mp 2\pi (n - (M + m))}{n}}  $, and $n - (\mathcal M + m)\leq \mathcal M$ by definition.}  }
\end{itemize}
More explicitly, we can write these basis vectors as:
\begin{align*}
|v_0\rangle &= [1,\;1,\;....\; 1] \\
|v'_0\rangle &= [1,\;-1,\;1,....\; -1] \qquad \text{($n$ is even)} \\
|v_1^{\pm}\rangle &= [1,\; e^{\pm i\frac{2\pi}{n}},\; e^{\pm 2i\frac{2\pi}{n}},\;....\; e^{\pm (n-1)i\frac{2\pi}{n}}] \\
|v_2^{\pm}\rangle &= [1,\; e^{\pm i\frac{4\pi}{n}},\; e^{\pm 2i\frac{4\pi}{n}},\;....\; e^{\pm (n-1)i\frac{4\pi}{n}}]  \\
\vdots & 
\\
|v_{M}^{\pm}\rangle &= [1,\; e^{\pm i\frac{2\pi M}{n}},\; e^{\pm 2i\frac{2\pi M}{n}},\;....\; e^{\pm (n-1)i\frac{2\pi M}{n}}]  
\end{align*}
Here each index denotes a dipole moment amplitude parallel to the nanorods long-axis.
Notably this list contains $n$ basis vectors ($\pm$ are distinct), which can be seen to be orthogonal by inspection.  
As the basis vectors are orthogonal and their number is equal to the rank of the matrix $M$, the basis vectors span the complete, $n$-dimensional, response space of the oligomer.  
Furthermore, each basis vector belongs to a unique irreducible representation, and any eigenmode can only transform according to a single irreducible representation when excluding accidental degeneracies.
Therefore these basis vectors are also necessarily the $n$ eigenmodes of our system.  
The corresponding eigenvalues follow from inserting these field profiles into the coupled dipole equations (\ref{eq:couplingM}), but eigenmodes of complex conjugate irreducible representations, {\it i.e.} eigenmodes separated only by $\pm$ phase dependency, are known to be degenerate in any reciprocal system~\cite{Hopkins2016, Liu2017}.    
Note that for both the circularly polarized plane wave or {\sl dark-field} illumination,  a coordinate system rotations of $\frac{2\pi}{n}$ is equivalent to a global phase shift of $\pm \frac{2\pi}{n}$.
This implies the response of each nanorod is equal to that of every other nanorod in the oligomer, except for a phase difference specified by global rotation angle between the two given nanorods: the profile of induced dipole moments has equal magnitude dipole moments and a phase profile from 0 to $\pm2\pi$ in $\frac{2\pi}{n}$ increments. 
This notably matches  $|v_1^{\pm}\rangle$, which is orthogonal to every other eigenmode of the system.  
A LCP or RCP dark-field thereby excites exactly the same eigenmode as an LCP or RCP plane wave.

\subsubsection*{Chiral signal in extinction}
The eigenmode we consider will have an excitation amplitude $a_0$ due to an external illuminating field $ \mathbf{E_0}$  given by unconjugated dot products $a_0=\int  \boldsymbol v  \cdot \mathbf{E_0} \,\mathrm{d}V$.
This follows from $\mathbf{\bbar M}$ in (\ref{eq:couplingM}), or $\mathbf{\bbar G}$ in (\ref{eq:G0_1}), being a complex symmetric operator, {\it i.e. } $\mathbf{\bbar M}= \mathbf{\bbar M}^T$,\;$\mathbf{\bbar G}(\mathbf r, \mathbf r')= \mathbf{\bbar G}(\mathbf r', \mathbf r)= \mathbf{\bbar G}(\mathbf r, \mathbf r')^T$.
We can therefore define the excitation field provided by our circularly polarized illumination as $\mathbf {E_0} = \lambda \boldsymbol{v}$, and the extinction $\sigma_{ext}$ experienced by $\mathbf{E_0}$ will then be given by: \mbox{$\sigma_{ext} = \frac{1}{2}\int \mathbf{E_0}^* \cdot \mathbf{J} \mathrm{d}V  = \frac{|a_0|^2}{2}\mathrm{Re}\{\lambda\}$}, assuming we use normalized eigenmodes.
Subsequently the chiral ratio for extinction $\Delta_{ext}=\frac{\sigma_{\mathrm{LCP}}-\sigma_{\mathrm{RCP}}}{\sigma_{\mathrm{LCP}}+\sigma_{\mathrm{RCP}}}$  is equal to the overlap between RCP and LCP dark fields and the given oligomer  $\Delta_{E}=\frac{|{a_0}|^2_{\text{\tiny LCP}}-|{a_0}|^2_{\text{\tiny RCP}}}{|{a_0}|^2_{\text{\tiny LCP}}+|{a_0}|^2_{\text{\tiny RCP}}}=\Delta_{ext}$,  because the eigenvalues cancel between numerator and denominator of $\Delta_{ext}$. 
The only avenue for circular dichroism is therefore a difference in the magnitude of overlap between LCP and RCP illuminations and the given eigenmode, that is: $|a_0|$.
In the dipole model, we also know that each dipole in our oligomer experiences the same excitation, hence $a_0$ in $\Delta_E$ reduces to simplify the amplitude of the applied field parallel the orientation of to a single dipole.

\bibliography{chiral}